\documentclass[english]{article}
\usepackage[T1]{fontenc}
\usepackage[latin9]{inputenc}
\usepackage{color}
\usepackage{textcomp}
\usepackage{graphicx}
\usepackage{esint}

\makeatletter
\newcommand{\lyxaddress}[1]{
\par {\raggedright #1
\vspace{1.4em}
\noindent\par}
}

\makeatother

\usepackage{babel}
\begin{document}

\title{Tolman's law in linear irreversible thermodynamics: a kinetic theory
approach}

\author{A. Sandoval-Villalbazo$^{1}$, A. L. Garcia-Perciante$^{2}$, D.
Brun-Battistini$^{3}$}

\maketitle

\lyxaddress{$^{1}$,$^{3}$ Depto. de Fisica y Matematicas, Universidad Iberoamericana,
Prolongacion Paseo de la Reforma 880, Mexico D. F. 01219, Mexico. }

\lyxaddress{$^{2}$Depto. de Matematicas Aplicadas y Sistemas, Universidad Autonoma
Metropolitana-Cuajimalpa, Artificios 40 Mexico D.F. 01120, Mexico.}
\begin{abstract}
In this paper it is shown that Tolman's law can be derived from relativistic
kinetic theory applied to a simple fluid in a BGK-like approximation.
Using this framework, it becomes clear that the contribution of the
gravitational field can be viewed as a cross effect that resembles
the so-called \emph{Thomson effect} in irreversible thermodynamics.
A proper generalization of Tolman's law in an inhomogeneous medium
is formally established based on these grounds.
\end{abstract}


\section{Introduction}

In 1930 Richard C. Tolman showed that thermal equilibrium can persist
within a temperature gradient provided that a gravitational field
is present, satisfying the relation: 
\begin{equation}
\frac{1}{T}\frac{dT}{dr}=-\frac{g}{c^{2}}\label{eq:Tolman1}
\end{equation}
which is often referred to as Tolman\textasciiacute{}s law \cite{Tolman}.
Later, in 1940, Eckart proposed that the heat flux in a special relativistic
fluid should be proportional to the temperature gradient and to the
\emph{hydrodynamic acceleration} of the fluid \cite{Eckart1-1}. The
corresponding constitutive equation, derived from phenomenological
arguments reads
\begin{equation}
J_{\left[Q\right]}^{\mu}=-h^{\mu\nu}\kappa_{T}\left(T_{,\nu}+\frac{T}{c^{2}}a_{\nu}\right)\label{eq:1}
\end{equation}
where $h^{\mu\nu}$ is the spatial projector given by
\begin{equation}
h^{\mu\nu}=g^{\mu\nu}+\frac{1}{c^{2}}\mathcal{U}^{\mu}\mathcal{U}^{\nu}\label{eq:2}
\end{equation}
with $\mathcal{U}^{\mu}$ being the hydrodynamic four-velocity, $\mathcal{U}^{\mu}\mathcal{U_{\mu}}=-c^{2}$
and $\kappa_{T}$ is the thermal conductivity. It seems attractive
to relate Eckart\textasciiacute{}s constitutive equation to Tolman\textasciiacute{}s
law since requiring a vanishing heat flux for equilibrium leads to
\begin{equation}
\frac{1}{T}\frac{\partial T}{\partial x^{\ell}}=-\frac{a_{\ell}}{c^{2}}\label{eq:3}
\end{equation}
which strongly resembles Eq. (\ref{eq:Tolman1}). However, in the
presence of stresses the hydrodynamic (macroscopic) acceleration is
not equal to gravity. Moreover, the coupling expressed by Eq. (\ref{eq:1})
does not relate a thermodynamic flux with a canonical force, since
$a_{\nu}$ is not the spatial gradient of a state variable as required
by irreversible thermodynamics. Also, this coupling has been proven
to be the cause of generic instabilities, first identified by Hiscock
and Lindblom in 1985 \cite{HL}\cite{GRG09}.

On the other hand, it is well known that kinetic theory provides a
framework in which constitutive equations can be established from
first principles, in complete consistency with linear irreversible
thermodynamics in the Navier-Stokes regime. In this paper, it is shown
that a non-vanishing contribution to the heat flux due to a classical
gravitational field arises within the framework of relativistic kinetic
theory. Such effect strongly resembles the \emph{Thomson effect} present
in plasmas under the influence of an electrostatic field. It becomes
equivalent to Tolman's law for a homogeneous medium at mild temperatures
in the presence of a gravitational field.

In section 2 the basic principles of general relativistic kinetic
theory within the BGK approximation are described. Section 3 is devoted
to the solution of Boltzmann's equation and to the formal expression
for the heat flux, in the presence of a gravitational field, using
the Chapman-Enskog expansion to first order in the gradients. In section
4, the corresponding transport coefficient is calculated showing that
Tolman's law follows by requiring a vanishing heat flux in a homogeneous
medium. A proper generalization of Tolman's law in an inhomogeneous
medium is also formally established. Conclusions and final remarks
are included in section 5.

\section{Relativistic kinetic theory}

The standard method for establishing transport equations as well as
constitutive relations for dissipative fluxes is well known, and has
been thoroughly discussed by several authors. The formalism has been
successful in many applications, including relativistic fluids \cite{ck,degroor,ChCow}.
In such program a collisional term for the kinetic equation must be
specified as a starting point. In a relaxation time approximation,
using the so-called BGK collision kernel for the relativistic gas
proposed by Marle \cite{ck}, the kinetic equation reads
\begin{equation}
v^{\alpha}f_{,\alpha}+a^{\mu}\frac{\partial f}{\partial v^{\mu}}=-\frac{f-f^{\left(0\right)}}{\tau},\label{eq:Boltzmann-1}
\end{equation}
where $f$ is the single particle distribution function, $f^{\left(0\right)}$
the local equilibrium (Juttner) distribution \cite{ck,degroor}, $v^{\mu}$
is the molecular four-velocity and $\tau$ is a parameter of the order
of the collisional time. Since molecules follow geodetic trajectories,
the molecular acceleration $a^{\mu}$ can be written in terms of the
curvature induced by a gravitational field as
\begin{equation}
a^{\mu}=-\Gamma_{\alpha\beta}^{\mu}v^{\alpha}v^{\beta}\label{eq:amu}
\end{equation}
Thus, Eq. (\ref{eq:Boltzmann-1}) becomes
\begin{equation}
v^{\alpha}f_{,\alpha}-\Gamma_{\alpha\beta}^{\mu}v^{\alpha}v^{\beta}\frac{\partial f}{\partial v^{\mu}}=-\frac{f-f^{\left(0\right)}}{\tau},\label{eq:Boltzmann}
\end{equation}
According to the Chapman-Enskog method, it is possible to establish
a formal solution to the Boltzmann equation in the Navier-Stokes regime
that reads: 
\begin{equation}
f=f^{(0)}+f^{(1)}\label{eq:distribution function}
\end{equation}
where $f^{\left(1\right)}$ is a correction to local equilibrium to
first order in the gradients. Substituting expression (\ref{eq:distribution function})
in Eq. (\ref{eq:Boltzmann}), and retaining only first order terms
in the gradients leads to 

\begin{equation}
f^{\left(1\right)}=-\tau\left\{ v^{\alpha}f_{,\alpha}^{\left(0\right)}-\Gamma_{\alpha\beta}^{\mu}v^{\alpha}v^{\beta}\frac{\partial f^{\left(0\right)}}{\partial v^{\mu}}\right\} \label{eq:f1}
\end{equation}
Using the local equilibrium hypothesis, we can write the derivatives
of $f^{\left(0\right)}$ in Eq. (\ref{eq:f1}) as:

\begin{equation}
f_{,\alpha}^{\left(0\right)}=\frac{\partial f^{\left(0\right)}}{\partial T}T_{,\alpha}+\frac{\partial f^{\left(0\right)}}{\partial n}n_{,\alpha}+\frac{\partial f^{\left(0\right)}}{\partial\mathcal{U}^{\mu}}\mathcal{U}_{;\alpha}^{\mu}\label{eq:aa}
\end{equation}
where $T$ is the local temperature and $n$ is the local number density.
Recalling that the Juttner function reads:

\begin{equation}
f^{(0)}=\frac{n}{4\pi c^{3}z\mathcal{K}_{2}\left(\frac{1}{z}\right)}\exp\left(\frac{\mathcal{U}^{\beta}v_{\beta}}{zc^{2}}\right)\label{eq:Juttnerf0}
\end{equation}
a direct calculation leads to

\begin{equation}
f_{,\alpha}^{\left(0\right)}=f^{\left(0\right)}\left\{ \frac{n_{,\alpha}}{n}-\left[1-\frac{\gamma}{z}+\frac{1}{2z}\left(\frac{K_{1}\left(\frac{1}{z}\right)}{K_{2}\left(\frac{1}{z}\right)}+\mathcal{G}\left(\frac{1}{z}\right)\right)\right]\frac{T_{,\alpha}}{T}+\frac{v_{\mu}}{zc^{2}}\mathcal{U}_{;\alpha}^{\mu}\right\} \label{eq:Juttnerexpanded}
\end{equation}
where $K_{n}\left(\frac{1}{z}\right)$ is the $n_{th}$ modified Bessel
function of the second kind, $\mathcal{G}\left(\frac{1}{z}\right)=\frac{K_{3}(\frac{1}{z})}{K_{2}(\frac{1}{z})}$
and $\gamma=\gamma_{(k)}$ is the standard Lorentz factor for the
molecules' chaotic velocity $k^{\ell}$ (for more details see Ref.\cite{ChCow}). 

Since the molecular four-velocity that appears in Eqs. (\ref{eq:Juttnerf0}-\ref{eq:Juttnerexpanded})
can be written as

\begin{equation}
v^{\alpha}=v_{\mu}h^{\mu\alpha}+\gamma_{\left(k\right)}\mathcal{U}^{\alpha}\label{eq:vmol}
\end{equation}
the first term of Eq. (\ref{eq:f1}) can be expressed as

\begin{equation}
v^{\alpha}f_{,\alpha}^{\left(0\right)}=v_{\mu}h^{\mu\alpha}f_{,\alpha}^{\left(0\right)}+\gamma_{\left(k\right)}\mathcal{U}^{\alpha}f_{,\alpha}^{\left(0\right)}\label{eq:valfaf}
\end{equation}
or
\begin{eqnarray}
v^{\alpha}f_{,\alpha}^{\left(0\right)} & = & f^{\left(0\right)}v_{\mu}h^{\mu\alpha}\left\{ \frac{n_{,\alpha}}{n}-\left[1-\frac{\gamma}{z}+\frac{1}{2z}\left(\frac{K_{1}\left(\frac{1}{z}\right)}{K_{2}\left(\frac{1}{z}\right)}+\mathcal{G}\left(\frac{1}{z}\right)\right)\right]\frac{T_{,\alpha}}{T}\right.\nonumber \\
 &  & \left.+\frac{v_{\mu}}{zc^{2}}\mathcal{U}_{;\alpha}^{\mu}\right\} +\gamma_{\left(k\right)}\left\{ \frac{1}{n}\left.\frac{\partial n}{\partial t}\right|_{CF}+\frac{v_{\mu}}{zc^{2}}\left.\frac{\partial\mathcal{U}^{\mu}}{\partial t}\right|_{CF}\right.\nonumber \\
 &  & \left.-\left[1-\frac{\gamma}{z}+\frac{1}{2z}\left(\frac{K_{1}\left(\frac{1}{z}\right)}{K_{2}\left(\frac{1}{z}\right)}+\mathcal{G}\left(\frac{1}{z}\right)\right)\right]\frac{1}{T}\left.\frac{\partial T}{\partial t}\right|_{CF}\right\} \label{eq:15}
\end{eqnarray}
where $\left.\frac{\partial}{\partial t}\right|_{CF}$ indicates a
derivative calculated in the comoving frame, where locally $\mathcal{U}^{\nu}=\left[0,c\right]$.
Following Hilbert's standard procedure \cite{Courant}, Euler equations
are now used in order to express the time derivatives in Eq. (\ref{eq:15})
in terms of the gradients. We recall that Euler's equations in the
comoving frame read (see Ref. \cite{ck}):

\begin{equation}
\left.\frac{\partial n}{\partial t}\right|_{CF}=-n\theta\label{eq:continuity comoving frame}
\end{equation}
\begin{equation}
\left.\frac{\partial\mathcal{U}^{\ell}}{\partial t}\right|_{CF}=-\frac{1}{\tilde{\rho}}\left[h^{\ell\nu}\left(\frac{T_{,\nu}}{T}+\frac{n_{,\nu}}{n}\right)-c^{2}\Gamma_{44}^{\ell}\right]\qquad\left.\frac{\partial\mathcal{U}^{4}}{\partial t}\right|_{CF}=0\label{eq:momentum comoving frame}
\end{equation}
\begin{equation}
\left.\frac{\partial T}{\partial t}\right|_{CF}=0\label{eq:energy comoving frame}
\end{equation}
In Eq. (\ref{eq:momentum comoving frame}) use has been made of the
fact that all four components of the projector $h^{\mu\nu}$ vanish
as well as $\Gamma_{44}^{4}$ in a static metric.

The last term of Eq. (\ref{eq:f1}) is easily established, and for
a newtonian metric reads:

\begin{equation}
a^{\mu}\frac{\partial f^{(0)}}{\partial v^{\mu}}=-\Gamma_{\alpha\beta}^{\mu}v^{\alpha}v^{\beta}\frac{\mathcal{U}_{\mu}}{zc^{2}}f^{(0)}\simeq-\Gamma_{44}^{\mu}\frac{\mathcal{U}_{\mu}}{z}f^{(0)}\label{eq:f(1) comoving frame}
\end{equation}
We are now in position to address the main task of this work namely,
to analyze the heat flux associated to a gravitational field applied
to a simple fluid.

\section{Heat flux}

In this section attention will only be paid to the gravitational terms
present in the first order in the gradients correction to the equilibrium
distribution function. Introducing Eqs. (\ref{eq:15}) to (\ref{eq:f(1) comoving frame})
in Eq. (\ref{eq:f1}) and ignoring all terms related to non-gravitational
potential gradients one obtains

\begin{equation}
f_{[g]}^{\left(1\right)}=-\tau\frac{1}{zc^{2}}f^{\left(0\right)}\left\{ \gamma_{\left(k\right)}v_{\mu}\frac{nm}{\tilde{\rho}}(-\Gamma_{44}^{\mu})+v_{\mu}(\Gamma_{44}^{\mu})\right\} ,\label{eq:f(1) cf external force}
\end{equation}
where $\tilde{\rho}=\frac{n\varepsilon}{c^{2}}+\frac{p}{c^{2}}.$
This expression clearly vanishes in the non-relativistic limit and
thus no heat flux is induced due to a gravitational field in a simple
non-relativistic fluid, as expected. However, in the relativistic
case and using the newtonian approximation for the metric, the following
expression can be written:
\[
v_{\mu}\Gamma_{44}^{\mu}\simeq\gamma_{\left(k\right)}k_{\ell}\frac{g^{\ell}}{c^{2}}
\]
and thus

\begin{equation}
f_{[g]}^{\left(1\right)}=-\tau f^{\left(0\right)}\gamma_{\left(k\right)}\frac{k_{\ell}g^{\ell}}{zc^{2}}\left\{ \frac{\gamma_{\left(k\right)}}{\mathcal{G}\left(\frac{1}{z}\right)}-1\right\} \label{eq:f(1) rel cf gravity}
\end{equation}

We now recall that the local heat flux in a relativistic fluid is
expressed in the Navier-Stokes regime as \cite{PhysA}:

\begin{equation}
J_{\left[Q\right]}^{\ell}=mc^{2}\int k^{\ell}f^{\left(1\right)}\gamma_{\left(k\right)}^{2}d^{*}K\label{eq:heatFlux rel}
\end{equation}
so that the thermal dissipation term that arises from a gradient in
the gravitational potential can be identified as:

\begin{equation}
J_{\left[Q,g\right]}^{\ell}=-\tau m\frac{1}{3}g^{\ell}\int f^{\left(0\right)}k^{2}\gamma_{\left(k\right)}^{3}\frac{1}{z}\left\{ \frac{\gamma_{\left(k\right)}}{\mathcal{G}\left(\frac{1}{z}\right)}-1\right\} d^{*}K,\label{eq:heatfluxgrav}
\end{equation}
Notice that the determinant of the metric is not included as a factor
in the integral since within this approximation deviations from a
unitary value are second order and higher in the gravitational potential
gradient. The invariant velocity volume element $d^{*}K$ is established
in Refs.\cite{Liboff,Ana y Alma}. Details on the procedure to evaluate
this type of integrals can also be found in these references. The
constitutive equation for the gravitational term in the heat flux
can then be written as

\begin{equation}
J_{\left[Q,g\right]}^{\ell}=-L_{Tg}g^{\ell}\label{eq:Jqg-2}
\end{equation}
where
\begin{equation}
L_{Tg}=\frac{nkT}{z}\tau\left\{ 5z+\frac{1}{\mathcal{G}\left(\frac{1}{z}\right)}-\mathcal{G}\left(\frac{1}{z}\right)\right\} ,\label{eq:LTg}
\end{equation}
The behavior of $L_{Tg}$ can be observed in Fig. 1 where $f_{g}\left(z\right)=L_{Tg}/(kT\tau n)$
is plotted as a function of $z$.
\begin{figure}
\caption{Dependence with $z$ of the coefficient that couples the heat flux
with the gravitational field. The coefficient tends to zero in the
small $z$, non-relativistic, limit and asymptotes to a constant value
for $z\gg1$ (ultra-relativistic systems).}

\begin{centering}
\includegraphics{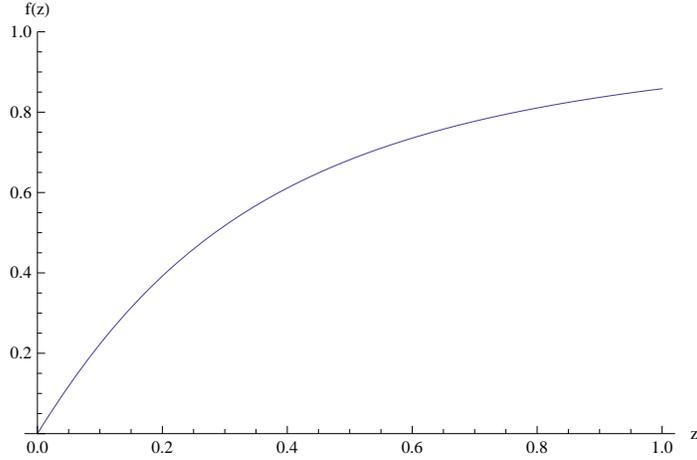}
\par\end{centering}

\label{fig:F1} 
\end{figure}

To the authors' knowledge this calculation is novel. In the next section
we shall analyze the low temperature limit of Eq. (\ref{eq:Jqg-2})
and relate the corresponding result to Tolman's law.

\section{Transport coefficients and Tolman's law}

For $z\ll1$, the expansion in Taylor series of the expression in
brackets in Eq. (\ref{eq:LTg}) reads:
\begin{equation}
5z+\frac{1}{\mathcal{G}\left(\frac{1}{z}\right)}-\mathcal{G}\left(\frac{1}{z}\right)\sim\frac{5}{2}\left(z^{2}-z^{3}+...\right)\label{eq:Taylor}
\end{equation}
and to lowest order in $z$ one obtains 

\begin{equation}
J_{\left[Q,g\right]}^{\ell}\sim-\frac{5}{2}\frac{nk^{2}T^{2}}{mc^{2}}\tau g^{\ell}\label{eq:Jqg z 2}
\end{equation}
If the fluid is assumed to be homogeneous (constant density), the
heat flux is expressed as follows:

\begin{equation}
J_{\left[Q\right]}^{\ell}=-L_{TT}\frac{T^{,\ell}}{T}-L_{Tg}g^{\ell},\label{eq:Jqg total inhomogeneous}
\end{equation}
where it is easily seen that since $L_{Tg}$ scales as $z$, the cross
effect that couples heat with gravity is only present in the relativistic
case. 

In order to relate this result with Tolman's law we assume a vanishing
heat flux as a requirement for equilibrium so that
\begin{equation}
\frac{T^{,\ell}}{T}=-\frac{L_{Tg}}{L_{TT}}g^{\ell}\label{eq:newTolman}
\end{equation}
Since the heat conductivity coefficient to lowest order in $z$ is
given by \cite{PhysA}

\begin{equation}
L_{TT}=\frac{5}{2}\frac{nk^{2}T^{2}}{m}\tau,\label{eq:Ltt}
\end{equation}
one finally obtains the standard form of Tolman's law:

\begin{equation}
\frac{T^{,\ell}}{T}=-\frac{1}{c^{2}}g^{\ell}\label{eq:Tolman recovered}
\end{equation}

It is worth to notice that for a non-homogeneous medium, the heat
flux expression involves three thermodynamic forces 

\begin{equation}
J_{\left[Q\right]}^{\ell}=-L_{TT}\frac{T^{,\ell}}{T}-L_{nT}\frac{n^{,l}}{n}-L_{Tg}g^{\ell}\label{eq:fullheatflux}
\end{equation}
Therefore in the general case of a simple relativistic non-degenerate
and inhomogeneous fluid, the following generalization of Tolman's
law is obtained:

\begin{equation}
L_{Tg}g^{\ell}=-L_{TT}\frac{T^{,\ell}}{T}-L_{nT}\frac{n^{,l}}{n}\label{eq:Tolman new}
\end{equation}
The complete expressions for the transport coefficients $L_{TT}$
and $L_{nT}$ have already been calculated in reference \cite{PhysA}
using the BGK approximation for the kernel in Boltzmann's equation.
Namely, 
\begin{eqnarray}
L_{TT} & = & \frac{nk^{2}T^{2}\tau}{m}\left\{ \left(\frac{1}{z}-\left(4z+\frac{K_{1}(\frac{1}{z})}{K_{2}(\frac{1}{z})}\right)^{-1}\right)\left(\frac{1}{z}+5\mathcal{G}\left(\frac{1}{z}\right)\right)\right.\nonumber \\
 &  & \left.-\left(1+\frac{K_{1}(\frac{1}{z})}{2zK_{2}(\frac{1}{z})}+\frac{\mathcal{G}\left(\frac{1}{z}\right)}{2z}\right)\frac{\mathcal{G}\left(\frac{1}{z}\right)}{z}\right\} \label{eq:LTT}
\end{eqnarray}

\begin{equation}
L_{nT}=\frac{nk^{2}T^{2}\tau}{m}\left\{ \frac{\mathcal{G}\left(\frac{1}{z}\right)}{z}-\left(4z+\frac{K_{1}(\frac{1}{z})}{K_{2}(\frac{1}{z})}\right)^{-1}\left(\frac{1}{z}+5\mathcal{G}\left(\frac{1}{z}\right)\right)\right\} ,\label{eq:LnT}
\end{equation}

It is important to point out that, in the case of an inhomogeneous,
non-isothermal fluid in the presence of a gravitational field, the
contribution for the total heat flux is of order zero in $z$ for
the temperature gradient and of first order in $z$ for the density
and gravitational potential gradients. Therefore, the temperature
gradient makes the most important contribution in the heat flux generation.

\section{Discussion and final remarks}

It is well known that a heat flux arises when a temperature gradient
is present in a simple fluid. Richard C. Tolman showed, based only
on phenomenological arguments, that a gravitational potential can
balance a temperature gradient in order to achieve thermal equilibrium.
In this paper it has been shown, using relativistic kinetic theory
arguments that, in a relativistic inhomogenous simple fluid, a heat
flux can be generated not only by a temperature and/or density gradient
but also by a gravitational field. This is expressed by Eq.(\ref{eq:fullheatflux})
and the corresponding gravitational transport coefficient has been
established in the Navier-Stokes regime using the first order Chapman-Enskog
method to solve Boltzmann's equation.

A further analysis of expression (\ref{eq:fullheatflux}), illustrates
the fact that the heat flux is only related to canonical forces. It
is important to notice that, in a continuous medium, the hydrodynamic
acceleration is not equal to the single particle acceleration. In
general Tolman's law should not be identified with Eq. (\ref{eq:3}). 

The expression for the total heat flux shows that two cross effects
are present in a \emph{single component relativistic fluid.} This
feature is completely absent in the single component non-relativistic
treatment in the presence of external fields. The present calculation
suggests that a non-vanishing diffusive flux may also exist in the
relativistic regime implying symmetry (Onsager's) relations for the
transport coefficients.

Kinetic theory provides a very powerful tool to explore the transport
properties of gases in the presence of strong gravitational field
barely studied in the past. The extension of this result to non-static
space-times is a very interesting subject which will be addressed
in the near future. 

\medskip{}

\textsf{\textbf{\textit{\textcolor{black}{\Large Acknowledgements}}}}{\Large \par}

The authors wish to thank L. S. Garcia-Colin and R. A. Sussman for
their valuable comments for this work and CONACYT for support through
grant number 167563.

\end{document}